
%
%

\input harvmac.tex

\input epsf.tex
\overfullrule=0mm
\newcount\figno
\figno=0
\def\fig#1#2#3{
\par\begingroup\parindent=0pt\leftskip=1cm\rightskip=1cm\parindent=0pt
\baselineskip=11pt
\global\advance\figno by 1
\midinsert
\epsfxsize=#3
\centerline{\epsfbox{#2}}
{\bf Fig. \the\figno:} #1\par
\endinsert\endgroup\par
}
\def\figlabel#1{\xdef#1{\the\figno}}
\def\encadremath#1{\vbox{\hrule\hbox{\vrule\kern8pt\vbox{\kern8pt
\hbox{$\displaystyle #1$}\kern8pt}
\kern8pt\vrule}\hrule}}
\overfullrule=0pt
%

%
\def\tilde{\widetilde}

\def\*{\star}
\def\[{\left[}
\def\]{\right]}
\def\({\left(}
\def\){\right)}
\def\frac#1#2{{#1 \over #2}}

\def\d{\partial}

\def\2pi{\hbox{$2\pi i$}}

\def\dsl{\raise.15ex\hbox{/}\kern-.57em\partial}
\def\Dsl{\,\raise.15ex\hbox{/}\mkern-.13.5mu D}


%
\Title{SPhT-96-139/ LAVAL-PHY-22/96}{\vbox {
\centerline{ Logarithmic Yangians in WZW models}   }}
\centerline{ D. Bernard$^\flat$\foot{Membre du C.N.R.S.},  Z. Maassarani$^\natural$
and P. Mathieu${^\natural}$ \foot{Work supported by NSERC (Canada) and FCAR
(Qu\'ebec).}
 }
\smallskip
\centerline 
\centerline{ $\qquad ^\natural$\it D\'epartement de
Physique, Universit\'e Laval, Qu\'ebec, Canada G1K 7P4}
\smallskip
{$\qquad ^\flat$\it Service de Physique Th\'eorique de Saclay, 
F-91191 Gif-sur-Yvette, 
France \foot{Laboratoire de la Direction des Sciences du Commisariat \`a
l'Energie Atomique.}}
\vskip.4in
\bigskip
\centerline{\bf Abstract}
\bigskip\vskip 0.45cm
 A new action of the Yangians in the WZW models is displayed. Its structure is
generic and level independent.   This Yangian is the natural extension at the
conformal point of  the one unravelled in massive theories with current algebras.
Expectingly, this new symmetry of WZW models will lead to a deeper 
understanding of the integrable structure of conformal field theories
and their deformations.

\Date{12/96}

\vfill\eject

\newcount\eqnum \eqnum=1
\def\eq{ \eqno(\secsym\the\meqno) \global\advance\meqno by1 }
\def\eqlabel#1{ {\xdef#1{\secsym\the\meqno}} \eq }

\newwrite\refs
\def\startreferences{ \immediate\openout\refs=references
 \immediate\write\refs{\baselineskip=14pt \parindent=16pt \parskip=2pt} }
\startreferences

\refno=0
\def\aref#1{\global\advance\refno by1
 \immediate\write\refs{\noexpand\item{\the\refno.}#1\hfil\par}}
\def\ref#1{\aref{#1}\the\refno}
\def\refname#1{\xdef#1{\the\refno}}

\let\y\infty

\let\rw\rightarrow

\let\l\left
\let\r\right
\let\R\rangle
\let\L\langle

\font\tenmib=cmmib10
\font\sevenmib=cmmib10 at 7pt
\font\fivemib=cmmib10 at 5pt
\newfam\mibfam 

\textfont\mibfam=\tenmib
\scriptfont\mibfam=\sevenmib
\scriptscriptfont\mibfam=\fivemib
\mathchardef\alphaB="080B
\mathchardef\betaB="080C
\mathchardef\gammaB="080D
\mathchardef\deltaB="080E
\mathchardef\epsilonB="080F
\mathchardef\zetaB="0810
\mathchardef\etaB="0811
\mathchardef\thetaB="0812
\mathchardef\iotaB="0813
\mathchardef\kappaB="0814
\mathchardef\lambdaB="0815
\mathchardef\muB="0816
\mathchardef\nuB="0817
\mathchardef\xiB="0818
\mathchardef\piB="0819
\mathchardef\rhoB="081A
\mathchardef\sigmaB="081B
\mathchardef\tauB="081C
\mathchardef\upsilonB="081D
\mathchardef\phiB="081E
\mathchardef\chiB="081F
\mathchardef\psiB="0820
\mathchardef\omegaB="0821
\mathchardef\varepsilonB="0822
\mathchardef\varthetaB="0823
\mathchardef\varpiB="0824
\mathchardef\varrhoB="0825
\mathchardef\varsigmaB="0826
\mathchardef\varphiB="0827


\newsec{Introduction}

In recent years  Yangian symmetry has been unraveled in various 
physical models. A particularly interesting manifestation
of this symmetry is its occurence in Haldane-Shastry [\ref{ F.D.M. Haldane, Phys.
Rev. Lett. {\bf 60} (1988) 635;  B.S. Shastry, Phys.
Rev. Lett. {\bf 60} (1988) 639.}] long-range interacting spin chains, 
which have been found  to realize a discrete analogue of level-1 WZW theories
[\ref{F.D.M. Haldane, Z.N.C. Ha, J.C. Talstra, D. Bernard
  and V. Pasquier, Phys. Rev. Lett. {\bf 69} (1992) 2021.}\refname\HH].  This
has led to a quasi-particle (spinon) description of these conformal models
[\ref{D. Bernard, V. Pasquier and D. Serban, 
  Nucl. Phys. {\bf B428} (1994) 612, (hep\-th/9404050).}\refname\BPS, \ref{K.
Schoutens,
  Phys. Lett. {\bf 331B} (1994) 335, (hep-th/9401154).}\refname\BLSa, \ref{P.
Bouwknegt, A.W.W. Ludwig and K. Schoutens,
  Phys. Lett. {\bf 338B} (1994) 448, (hep-th/9406020).}\refname\BLSb, \ref{ P.
Bouwknegt, A.W.W Ludwig and K. Schoutens,
  Phys. Lett. {\bf 359B} (1995) 304, (hep-th/9412108)}\refname\BLSc, \ref{ P.
Bouwknegt,
 and K. Schoutens, {\it The $\widehat{SU}(n)_1$ WZW
models}, (hep-th/9607064)}\refname\BLSd]  expected to be tailor-made for the
analysis of off-critical deformations.
Although rather explicit, this construction has some intrinsic limitations:
except for $su(N)$-WZW models at level 1, there are no explicit description of the
Yangian currents and the $su(N)$ results do not
generalize directly to other algebras (see e.g., [\BLSc]).

This situation contrasts sharply with
that pertaining to the integrable massive deformations of the CFT theories for
which there is a Yangian symmetry for any affine Lie spectrum generating algebra
at the fixed point, with a level-independent structure 
[{\ref{D. Bernard,Comm. Math,. Phys. {\bf 137} (1991)
191; cf also ``Quantum symmetries in 2D massive field theory", 
Cargese '92 proceedings,hep-th/9109058.}\refname\denis,
\ref{M. L\"uscher, Nucl. Phys.  {\bf B135} (1978) 1.}\refname\lush,
\ref{ H.J. de Vega, H. Eichenherr and J.M. Maillet, Nucl.
Phys. {\bf B240} (1984) 377; Comm. Math. Phys. {\bf 92} (1984) 507.}\refname\vema].  These
Yangian generators are nothing but the direct quantum extensions of the classical
nonlocal charges of the sigma model or current algebra models [\ref{M. L\"usher
and K. Polmeyer, Nucl. Phys. {\bf B137} (1978) 46; E. Brezin, C. Itzykson, J.
Zinn-Justin and J.B. Zuber, Phys. Lett. {\bf B 82} (1979) 442.}\refname\popol].

In the present work, we display a new Yangian symmetry in WZW models whose
structure is generic and level independent. It provides a direct
extension of the off-critical Yangian realization in massive theories. 
Unfortunately, what is gained in generality seems to be lost in explicitness.

\newsec{Nonlocal currents in WZW models}

The classical equations of motion for the  WZW models at the conformally
invariant points can be written as the 
conservation laws, $\partial_\mu J^\pm_\mu=0$,
for two chiral currents $J^\pm_\mu$, respectively defined by
$$J^+_\mu = g^{-1}\d_\mu g - \epsilon_{\mu\nu}g^{-1}\d_\nu g, \qquad\qquad 
J^-_\mu= \d_\mu g g^{-1} + \epsilon_{\mu\nu}\d_\nu g g^{-1}\eq$$
As a consequence, there exists classically two nonlocal conserved
currents defined by:
$$ {\cal J}_\mu^{\pm\, a}(x) = i f^{abc} \int^x dv^\sigma\epsilon^{\sigma\rho} 
J^{\pm\, b}_\rho(v) J_\mu^{\pm\, c}(x)\eqlabel\defpopol $$
Similarly, in most  nonlinear sigma models, the equations of motion
can be written as a conservation law for a curl-free current taking values in
a finite dimensional Lie algebra. This ensures the existence of nonlocal 
currents [\popol] given by an expression similar to (\defpopol). In 
favorable cases these currents persist in the quantum theory. They are then
the generators of a Yangian algebra [\lush,\denis,\vema].

The structure of the nonlocal conserved currents in massive theories suggests the
following form for the nonlocal density of the Yangian generator  in the ultraviolet
limit:
$$ Y^a(z; P) = \lim_{y\rw z}\l\{i f^{abc}\int_P^z dv J^b(v)J^c(y) + 2g \,J^a(y)\,
\ln \big({z-y\over P-y}\big) \r\} \eqlabel\ydef$$
where $J^a(z)$ are the generators of the spectrum generating current algebra,
$$J^a(z) J^b(w) \sim  {if^{abc}\, J^c(w) \over z-w } +{k\, \delta^{ab} \over
(z-w)^2}\eq$$
and $f^{abc}$ are the completely antisymmetric structure constants, normalized according to
$$ f^{abc} f^{abd} = 2g\, \delta^{cd}\eq$$ $g$ stands for the dual Coxeter number
and
$k$ is the level. The first integration contour in eq.(\ydef), $\int_P^z dv
J^b(v)$, is the same as the contour used to define the second logarithmic term,
$\ln(\frac{z-y}{z-P})=\int_P^z \frac{dv}{v-y}$.
Note that (\ydef) could equally well be written as
$$ Y^a(z; P) = {1\over 2\pi i} \oint {dx\over x-z}\l\{i
f^{abc}\int_P^x dvJ^b(v)J^c(z) + 2g \,J^a(z)\,
\ln \big({x-z\over P-z}\big)\r\} \eqlabel\yydef$$
which is the usual way field products are regularized in conformal theories.

The main observation that needs to be done from the expression of this nonlocal current
is the dependence upon a certain point $P$.  This is a novelty brought by the
conformal invariance which prevents us to simply let $P\rw \y$ from the
onset.  Much of the complications to follow are rooted in this simple fact.   In
the massive regime,
$P$ can be sent to $\y$ since correlation functions decrease exponentially;
mathematically said,  the points are no longer all conformally equivalent.

It should also be mentionned that this candidate nonlocal current is not unique.  The
following variant of  $Y^a$ is also regular and shares most of its properties:
$$ {\tilde Y}^a(z; P) = \lim_{y\rw z} \l[i  f^{abc}\int_P^z dvJ^b(v)J^c(y) +
2g\,J^a(y)\ln (z-y)\r]\eqlabel\ytdef$$ 

In view of relating the charges associated to these currents with Yangian
generators, we 
first need to evaluate the effect of inserting nonlocal currents in correlation
functions, that is, to calculate 
$\L Y^a(z;P)\prod_j \phi_j(\zeta_j)\R$.  Here  $\phi_j(\zeta_j) $ stands for a WZW
primary field:
$$J^a(z) \phi_j(w) \sim {-t^a_j\over z-w} \phi(w)\eq$$ where $t^a_j$ is a matrix
representation (specified  by $j$) of the symmetry algebra: $[t^a_j,t^b_k] =
i\delta_{jk}f^{abc}t^c_j$. A direct calculation yields
$$\L Y^a(z;P)\prod_j \phi_j(\zeta_j)\R = i f^{abc} \sum_{j,k} {t^b_jt^c_k\over
z-\zeta_k} \ln \big({z-\zeta_j\over P-\zeta_j}\big)\L \prod_j
\phi_j(\zeta_j)\R\eq$$ The nonlocal structure of $Y^a$ manifests itself through 
logarithmic terms. Once again, the logarithmic terms $\ln(\frac{z-\zeta_j}{P-\zeta_j})$
are defined by the integrals $\int\frac{dv}{v-\zeta_j}$ where the integration
contours are paths from $P$ to $z$. The nonlocality of $Y(z;P)$ is
encoded in the monodromy acquired by  these integrals upon
 contour deformations.

\newsec{ Nonlocal charges in WZW models and comultiplication}

The action of the charge $Q_1^a$ -- associated to the current $Y^a(z;P)$ --  on a field
$\phi_0(\zeta_0)$ is defined as follows [\denis]:
$$\L Q^a_1(\phi_0(\zeta_0)) \prod_{j\geq 1} \phi_j(\zeta_j)\R  =
\oint_{\gamma_0}dz \L Y^a(z;P)\prod_{j\geq 0}
\phi_j(\zeta_j)\R\eqlabel\action$$
where the contour $\gamma_0$ is defined as in Fig. (1a).  
Essentially, the contour
circles around $\zeta_0$ but excludes all other $\zeta_i$'s. However, the presence of
the logarithmic cut calls for a more careful specification.  The cut, which extends from
$\zeta_0$ to infinity is chosen to pass through the point $P$ and the contour
includes $P$ although it remains slightly open due to the cut. The point $z$
starts then just below (and at the left of) $P$, hence just below the cut, 
follows the cut up to $\zeta_0$ which is then encircled and
returns to $P$ above the cut (and again at the left of $P$). The phase of
$(z-\zeta_0)$  above the cut differs by
$2\pi i $ from its value just below it.\foot{For subsequent applications of the
charge $Q_1^a$, we mention that, in Fig. (1a), the open contour around $P$ 
has vanishing radius. Thus, for this part of $\gamma_0$, the logarithmic terms
do not contribute, while the pole contributions are picked up (which can 
arise for some  of the terms in a correlation function), {\it as if},
in the absence of cuts, the contour was completely closed at the left of $P$.}

In this way, making use of the first and third contour integrals of appendix A,
we find the explicit expression for the action of the charge $Q_1^a$
on a primary field:
$$
\L Q^a_1(\phi_0(\zeta_0)) \prod_{j\geq 1} \phi_j(\zeta_j)\R = 
 \l\{ {-gi\pi} t_0^a
+ 2i f^{abc} \sum_{k\not=0} {t^b_kt^c_0}  \ln \big({\zeta_0-\zeta_k\over
P-\zeta_k}\big)\r\}\L\phi_0(\zeta_0)
\prod_{j\geq 1}\phi_j(\zeta_j)\R\eq$$

On the other hand, the action of the local charge $Q_0^a$ on $\phi_0$ is defined
as usual in terms of a small contour around $\zeta_0$.  This results in 
$Q^a_0(\phi_0)=-t^a_0\phi_0$. We recall that $Q^a_0$ acts additively on
a product of fields.

The next step is to define the action of $Q_1^a$ on two fields $\phi_1(\zeta_1)$
and $\phi_2(\zeta_2)$.  This is still defined by a contour of integration:
$$\L Q^a_1(\phi_1(\zeta_1)\phi_2(\zeta_2)) \prod_{j\geq 3} \phi_j(\zeta_j)\R  =
\oint_{\Gamma} dz \L Y^a(z;P)\prod_{j\geq 0}
\phi_j(\zeta_j)\R\eq$$
but with a different integration contour, which is illustrated in Fig. (1b). 
This contour can be deformed in
two contours $\gamma_1$ and $\gamma_2$. However, after the first contour a shift of
$2\pi i$  occurs, which is responsible for the following comultiplication
rule
$$\Delta  Q^a_1 =  Q^a_1\otimes 1+ 1\otimes  Q^a_1 -2\pi f^{abc} Q_0^b\otimes  Q_0^c
\eq$$ This is indeed the comultiplication of Yangians.

\newsec{Yangian relations}

We will now show that the charges $Q_0^a$ and $Q_1^a$ are the first two
generators of a Yangian symmetry, ie. that they generate a representation
of the Yangian algebra. This is our main result.
It  amounts to establishing the relations
$$\eqalign{
&{\rm Y(1):}\quad  [Q_0^a, Q_0^b] = if^{abc} Q_0^c\cr
&{\rm Y(2):}\quad  [Q_0^a, Q_1^b] = if^{abc} Q_1^c\cr
& {\rm Y(3):}\quad  [Q_1^a, [Q_1^b, Q_0^c]] - [Q_0^a, [Q_1^b, Q_1^c]]=
C\,A^{abc;def} \{Q_0^d,Q_0^e,Q_0^f\}_s\cr}\eqlabel\yang$$
where 
$$A^{abc;def} =  f^{adp}f^{beq}f^{cfr}f^{pqr}\eq$$
 $\{ a,b,c\}_s$ stand for
the complete symmetrization of the three objects
 $a,b,c$ (which incorporates a factor
$1/6$) and $C$ is a constant fixed by the comultiplication.
 With our normalization for
the comultiplication, we have
$$C=-4\pi^2\eq$$ A fourth relation, needed only for $su(2)$, will be ignored
here. 

We will prove the Yangian relations (\yang) by checking them for the action
of the charges $Q^a_0$ and $Q^a_1$ on a set of primary fields. 
(Actually, this ckeck provides only a partial proof of the Yangian relations
since acting with the Yangian generators produces fields which are not
generated by products of primary fields, i.e., nonlocal fields are
produced.) As is well known, the action of $Q_0^a$ obtained previously  implies 
Y(1) directly. Having obtained the Yangian comultiplication, it is enough to
verify the basic relation Y(2) and the
Serre relation Y(3) on a single field,  since this will imply
that they are also verify on an arbitrary number of primary fields.
Hence, it suffices to evaluate the following expectation values
$$\L \l [Q_i^a(Q_1^b (\phi_0))-Q_1^b(Q_i^a (\phi_0))\r](\zeta_0)  
\prod_{j\geq 1}\phi_j(\zeta_j)\R \eq$$ 
with $i=0,1$.  Let us start with Y(2):
$$ \eqalign{
&\qquad\qquad\qquad\qquad\L Y^a(z;P)J^b(w)\prod_j \phi_j(\zeta_j)\R = \cr
&= \l[ - f^{adc} f^{dbe} \ln \big({z-w\over P-w}\big) 
\l\{ {k \delta^{ce}\over (z-w)^2}
+\sum_k{-i f^{cep}t^p_k \over (z-w) (w-\zeta_k)} 
\quad + \sum_{k,\ell}{t^c_kt^e_\ell \over (z-\zeta_k)
(w-\zeta_\ell)}\r\} \r.\cr
&  -i f^{adc}\sum_n t^d_n \ln \big({z-\zeta_n \over P-\zeta_n}\big) 
\l\{ {k\delta^{cb}\over (z-w)^2} + \sum_k {-i f^{cbp}t^p_k \over (z-w) (w-\zeta_k)} 
+\sum_{k,\ell} {t^c_kt^b_\ell \over (z-\zeta_k) (w-\zeta_\ell)}\r\} \cr
&\quad \l. +ikf^{abc} \l({1\over z-w } - {1\over P-w }\r) 
\sum_k {t^c_k\over z-\zeta_k} \r] \L\prod_j \phi_j(\zeta_j)\R\cr}\eq$$ 
The
commutator Y(2) is calculated from the difference between 
the two integration contours (see Fig. 2):
$$\oint_{\gamma_0} dz \oint_{\zeta_0} dw - 
\oint_{\gamma_0} dw \oint_{\gamma_0 } dz\eqlabel\cont$$
In the first step we integrate $w$ around $\zeta_0$ and then integrate $z$
along $\gamma_0$. The order of integration is reversed in the second step. Since
there are logarithmic cuts, $z$ must still be integrated along $\gamma_0$. 
The second integration contour encloses the first one: $w$
follows then a contour similar to 
$\gamma_0$. However, we stress that in absence of cuts (and this indeed arises
for some terms which  display only a pole structure), the small
contour around $P$ picks up pole contributions there.
After performing all integrals, the resulting commutator is found to be equivalent
to the action of $Q_1^a$ on the field
$\phi_0$, i.e., Y(2) is satisfied.\foot{Note that this requires the cancellation
of a term proportional to $\ln (1) - i\pi$, which means that a phase choice for
$\ln (-1)$ is required.}

The first step of the rather long calculation leading to Y(3) is to obtain
$$\L Y^a(z;P)Y^b(w;P)\prod_j \phi_j(\zeta_j)\R\eq$$whose appropriate integrations
will produce the commutator $[Q_1^a, Q_1^b]$.
An immediate difficulty must be bypassed which is that the
resulting correlation function is singular simply because the same point $P$
appears in the two currents.  This calls for a point-splitting regularisation:
$$\L Y^a(z;P)Y^b(w;P)\prod_j \phi_j(\zeta_j)\R= \lim_{P'\rw P}\L
Y^a(z;P')Y^b(w;P)\prod_j \phi_j(\zeta_j)\R\eq$$ 
Once the above correlator is evaluated, the integrals are
 performed along the following
contours (see Fig. 1):
$$\oint_{\gamma_0\cup\gamma_{P}} dz \oint_{\gamma_0} dw - \oint_{\gamma_0\cup\gamma_{P'}} dw
\oint_{\gamma_0 } dz\eq$$
where $\gamma_P$ is defined like $\gamma_0$.
Some computational simplifying features should be noticed. 
First, symmetrizing with respect to $P$ and $P'$ in intermediate calculations 
does not affect the final result and provides substantial simplifications.
Second, for the present purpose the commutator $[Q_1^a, Q_1^b]$ does not need to be
evaluated {\it in extenso} because not all resulting terms contribute to Y(3).  For
instance, it is simple to see that those terms proportional to $f^{abc}{\cal F}^c$ for some
functions
${\cal F}^c$ do not contribute and can thus be ignored.
Performing all the integrals (a collection of which is displayed in the Appendix) and
substituting the results into 
$$if^{bcd}[Q_1^a, Q_1^d]+ if^{abd}[Q_1^c, Q_1^d]+if^{cad}[Q_1^b, Q_1^d]\eq$$
(the left hand
side of Y(3)), we indeed recover the right hand side of Y(3), including the correct
normalizing factor $C$.\foot{For this calculation, there is no phase ambiguity
since $\ln(-1)$ appears only squared.} 
Thus $Q_0^a$ and $Q_1^a$ have been proved to be the first two Yangian generators.
 
A  quartic relation, Y(4), exists for all algebras; however it is a 
consequence of Y(2) and Y(3) for all algebras other than $su(2)$ 
[\ref{V.G. Drinfel'd, Sov. Math. Dokl. {\bf 32} (1985) 254.}]. 
For $su(2)$, the relation  Y(3) is trivial (both sides vanish) and 
the quartic relation has to be verified. We have not carried out this long
calculation but we believe that Y(4) should be satisfied in a  manner
similar to Y(3).    

\newsec{Yangians densities as logarithmic operators}

The OPE of the energy-momentum tensor and the nonlocal current is easily found to be:
$$ T(w) Y^a(z;P) \sim {  Y^a(z;P)\over (w-z)^2} + {\partial_z  Y^a(z;P)\over w-z}
- { J^a(z)\over (w-z)^2} + {  J^a(z)\over (w-z)(w-P)}-{G^a(z;P)\over w-P}\eq$$ 
where $ G^a(z;P)= i f^{abc} J^b(P)J^c(z)$.
Inserted in a correlation, this can be used to evaluate the commutator
$[L_{-1}, Q_1^a]$ acting on the field $\phi_0(\zeta_0)$ using the contours
(\cont) (see Fig. 2). The result is simply
$$ \l[\oint_{\gamma_0} dz
\oint_{\zeta_0} dw -
\oint_{\gamma_0} dw \oint_{\gamma_0 }dz\r]  \L Y^a(z;P) T(w)\prod_j
\phi_j(\zeta_j)\R = -2if^{abc}\sum_{k\not= 0} {t^b_0t^c_k\over
P-\zeta_k}\L \prod_j
\phi_j(\zeta_j)\R\eq$$
which vanishes only when $P\rw \y$.
Thus, if $Q^a_1\vert_\infty$ refers to the charge associated to
$ Y^a(z;\infty)$ we have:
$$ \[ L_{-1}, Q^a_1 \]\vert_\infty = 0 \eq$$
This commutation relation has to be used carefully since it holds
only in the limit $P\to\infty$ and with the limit taken after having performed 
the integration.
The fact that the Yangian charges commute with the momentum operator
$L_{-1}$ only once the limit $P\to\infty$ has been taken is the analogue of 
the well known fact that  the  quantum group symmetry
of integrable lattice models only emerges once the infinite lattice
limit as been taken.

When using this commutation relation, one has to remember that taking 
the limit $P\to\infty$ and integrating along the contour $\gamma_0$
are operations that do not commute.  This noncommutativity of the limits
prevents us from writting down directly the Ward indentities for the charges
$Q^1_a\vert_\infty$. A similar phenomenon for nonlocal symmetry on finite lattices
has been pointed out in [\ref{D. Bernard, G. Felder, Nucl. Phys. {\bf B365} (1991) 98.}].

If the point $P$ could be set to $\y$, the OPE of $T$ with $Y^a(z;\infty)$ 
would simply be
$$ T(w) Y^a(z;\infty) \sim {  Y^a(z;\infty)\over (w-z)^2} 
+ {\partial_z  Y^a(z;\infty)\over w-z}- { J^a(z)\over (w-z)^2} \eq$$
It reveals the
standard Jordan cell structure observed in [\ref{V. Gurarie, Nucl. Phys.
{\bf B410} (1993) 535.}], which
characterizes the presence of logarithmics operators. More precisely, it arises
if  at least two operators with the same dimension appear in the OPE of two other 
operators. Here the nonlocal operator $Y^a(z;\infty)$ would be the logarithmic
partner of the operator $J^a(z)$. More generally by acting recursively on a 
local field of the WZW model with $Y^a(z;\infty)$ one will produce an infinite
tower nonlocal fields [\denis, \ref{F. Smirnov, Int. J. Math. A7 (suppl) B1
(1992) 813.}]   which  are recursively logarithmic partners and which
form a Yangian representation. 
But once again one has to remember that the limit $P\to\infty$ does not
commute with the contour integrals.

\newsec{Conclusions}

We have displayed a new Yangian symmetry in WZW models that extends to the
massless regime the one  already observed in masssive theories 
with current algebras
and in sigma models. It holds for any affine Lie spectrum generating algebra and
any value of the level.  It should be stressed that  
this new Yangian symmetry  is rather different
from the one associated to the  spinon description
[\HH,\BPS,\BLSa,\BLSb,\BLSc,\BLSd]. Indeed, the spinon description provides a basis
of the WZW model Hilbert space  which diagonalizes an infinite set of commuting
hamiltonians containing the boost operator $L_0$.  The Yangian for which the
$n$-spinon states form irreducible multiplets therefore commutes
with these hamiltonians and thus with $L_0$, but not with the
momentum operator $L_{-1}$. On the contrary the
Yangian symmetry displayed here is more closely related
to the $S$-matrix description of conformal field 
theory [\ref{A.B. Zamolodchikov,
Al. B. Zamolodchikov, Nucl. Phys. {\bf B 379} (1992) 602.}]. For the WZW models, the
$S$-matrix is Yangian invariant. The Yangian multiplets are then formed
by the $n$-asymptotic particle states which are eigenstates of
an infinite set of hamiltonian containing the momentum operator $L_{-1}$. 
This provides an on-shell definition of the Yangian symmetry commuting
with $L_{-1}$ but not with $L_0$.
We have presented the off-shell construction of this
Yangian symmetry. We have seen that the commutation relations with the
boost and momentum operator are more subtle than for the on-shell
Yangian generators.
 The Yangian action we described and the one associated to the
spinon description are also distinguished by the way they act
on product fields. The logarithmic Yangian acts by comultiplication, 
which is not invariant under field permutations, whereas the action of the
spinon Yangian is invariant upon a permutation of the fields.\foot{That the
Yangian does not act by comultiplication on a $n$-spinon state can be seen
explicitly from eq.(4.6) in [\BPS]; it follows from  the presence of the
$\theta_{ij}$ factor in the action of $Q_1^a$.}  However,
both are compatible with the product operator expansions.

 To undertsand better the representations  associated to
the action of the Yangian charges on the fields
clearly remains an open problem.
Since the logarithmic Yangian acts on-shell on the $n$-particle states,
to introduce these representations will be necessary if one wants
to preserve the particle-field duality for massless theories.
It would be interesting to see whether this action could be intertwined
by transfer matrices associated to inhomogeneous spin chains.
Another problem is to write the Yangian Ward identities in a compact form, the
difficulty being related to coutour deformations in presence of
logarithmic cuts. Note finally that in view of the S-matrices of the
massive (current-current perturbations of WZW models)
theories, which are products of Yangian matrices and R-matrices of the RSOS
type, the logarithmic Yangian is expected to commute with the generators of the
(restricted) affine quantum group symmetries
${\cal U}_q({\cal G})$ which are associated to the RSOS factor [\ref{D.Bernard
and A. Leclair, Phys. Lett. {\bf 247B} (1990) 309.}].

\appendix{A}{Useful integrals and formulas}

Here we give a short list of useful integrals involving logs -- $f(z)$ stands for a
function which is  analytic everywhere on and inside the contour $\gamma$:
$$\eqalign{ 
& {1\over 2\pi i} \oint_{\gamma_\zeta} {dw} f(w)\, \ln\big({w-\zeta\over
P-\zeta}\big) =  \int_\zeta^P  dw\,{f(w)}\cr
& {1\over 2\pi i} \oint_{\gamma_\zeta} {dw} f(w)\, \ln^2\big({w-\zeta\over
P-\zeta}\big) =  2\int_\zeta^P  dw\,f(w) \l[i \pi   
+\ln\big({w-\zeta\over P-\zeta}\big)\r] \cr
& {1\over 2\pi i} \oint_{\gamma_\zeta} {dw\over w-\zeta} f(w)\, \ln\big({w-\zeta\over
P-\zeta}\big) = i\pi f(\zeta) + \int_\zeta^P  dw{f(w)-f(\zeta)\over w-\zeta}\cr
& {1\over 2\pi i} \oint_{\gamma_\zeta} {dw\over w-\zeta} \, \ln^2\big({w-\zeta\over
P-\zeta}\big) = -{4\pi^2\over 3} \cr
& {1\over 2\pi i} \oint_{\gamma_\zeta} {dw\over (w-\zeta)^2} \, \ln\big({w-\zeta\over
P-\zeta}\big) = -{1\over P-\zeta} \cr}\eq$$
To this list we add the defining relation of the dilogarithm function $Li_2(x)$:
$$Li_2(x) = -\int_0^x dx {\ln (1-x)\over x}\eq$$
which allows us to express the remaining log integrals, i.e.,
$$ \int_{\zeta}^{P'} {dw\over w-\zeta} \ln \big({P-w\over P-\zeta}\big) =
-Li_2\big({P'-\zeta\over P-\zeta}\big)
\eq$$
When integrating a dilogarithm function, we need to be careful about the cuts; this
function has a cut from 1 to $\y$. To avoid the two troublesome points, we can always
transform the argument by means of one of the following two relations [\ref{L. Lewin,
{\it Polylogarithms and associated functions}, North Holland, 1981.}]:
$$\eqalign{
&Li_2(-1/x) +Li_2(-x) = -{\pi^2\over 6} -{1\over 2} \ln^2 x\cr
&Li_2(x) +Li_2(1-x) = {\pi^2\over 6} -\ln x \ln(1-x)\cr}\eq$$


\bigskip \hrule \bigskip \centerline{{\bf References}}

\immediate\closeout\refs \vskip 0.5cm
  \message{References}\input references

\fig{a) The contour $\gamma_0$ consists of an open contour of vanishing 
radius, starting below and to the left of $P$, and of a contour encircling
$\zeta_0$ (dashed line). The second contour can be deformed into the contour
drawn with a solid line. b) The contour (broken line)
circling around $\zeta_1$ and
$\zeta_2$ is deformed into the contours drawn with solid lines.}{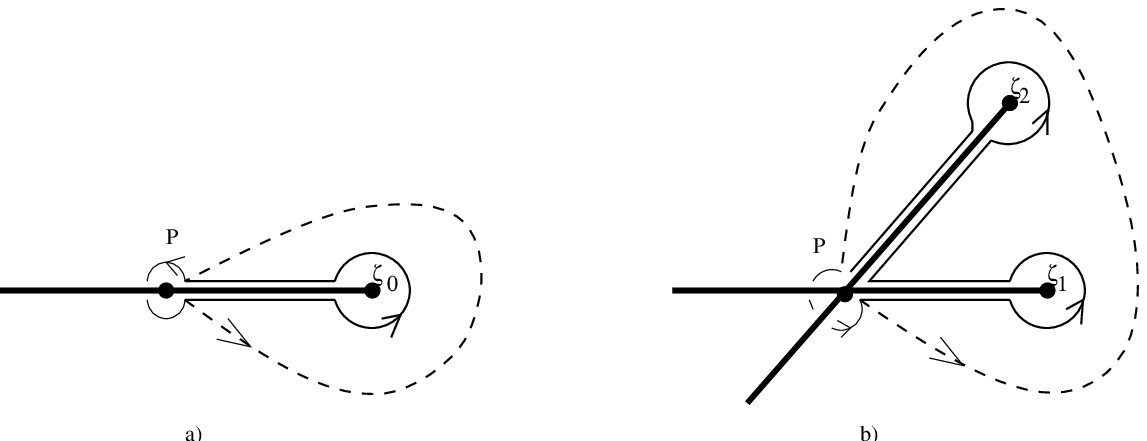}{15cm}
\figlabel\cont

\fig{For the first term, we carry out the $w$-integration  on the inner circle
first  and
then the $z$-integration  on the outer contour. For the second term 
the $z$-integration  is carried out on the inner contour
and the $w$-integration  on the outer contour.}{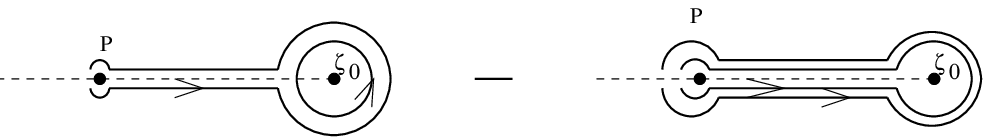}{15cm}
\figlabel\contA

\end